\documentclass[onecolumn,aps,prd,nofootinbib,showpacs,twoside,notitlepage,superscriptaddress]{revtex4-2}

\usepackage{amsmath,bm}
\usepackage{hyperref}
\usepackage{enumerate}
\usepackage{graphicx}
\usepackage{bm}
\usepackage{hyperref}
\usepackage{graphicx}
\usepackage[dvipsnames]{xcolor}
\usepackage{amsfonts}
\usepackage{amssymb,amsthm}
\usepackage{mathrsfs}
\usepackage{bbold}
\usepackage{orcidlink}

\newcommand{\ham}{\mathcal{H}}
\newcommand{\diff}{\mathcal{D}}
\newcommand{\shift}{{N^x}}
\newcommand{\lapse}{N}
\newcommand{\erad}{E^x{}}
\newcommand{\ephi}{E^\varphi{}}
\newcommand{\krad}{K_x}

\newcommand{\ephinew}{{E}^\varphi{}}
\newcommand{\kangnew}{{K}_\varphi}
\newcommand{\pphi}{P_\phi}

\begin{document}

\title{Nonsingular collapse of a spherical dust cloud}

\author{Asier Alonso-Bardaji\,\orcidlink{0000-0002-8982-0237}\,}
  \email{asier.alonso@ehu.eus}
	\affiliation{
 	Aix Marseille Univ., Univ. de Toulon, CNRS, CPT, UMR 7332, 13288
Marseille, France}

\author{David Brizuela\,\orcidlink{0000-0002-8009-5518}\,}
  \email{david.brizuela@ehu.eus}
	\affiliation{
Department of Physics and EHU Quantum Center, University of the Basque Country UPV/EHU,
Barrio Sarriena s/n, 48940 Leioa, Spain}

\begin{abstract}
We provide a covariant framework to study singularity-free Lema\^itre-Tolman-Bondi spacetimes with effective corrections motivated by loop quantum gravity.
We show that, as in general relativity, physically reasonable energy distributions lead to a contraction
of the dust shells. However, quantum-gravity effects eventually stop the collapse,
the dust smoothly bounces back, and no gravitational singularity is generated.
This model is constructed by deforming the Hamiltonian constraint of general relativity
with the condition that the hypersurface deformation algebra is closed.
In addition, under the gauge transformations generated by the deformed constraints,
the structure function of the algebra changes adequately, so that
it can be interpreted as the inverse spatial metric. Therefore, the model is completely
covariant in the sense that gauge transformations in phase space simply correspond to coordinate
changes in spacetime. However, in the construction of the metric, we point out a specific
freedom of considering a conformal factor, which we use to obtain a family of singularity-free spacetimes
associated with the modified model.
\end{abstract}

\maketitle

\section{Introduction}

Lema\^itre \cite{Lemaitre:1933gd} and Tolman \cite{Tolman:1934za} considered
a spherically symmetric dust (pressureless perfect fluid) to study the expansion of the Universe.
This model was also later studied by Bondi \cite{10.1093/mnras/107.5-6.410},
and it is usually referred to as the LTB model.
Although originally designed as a cosmological model in order to provide
a generalization of the homogeneous and spatially isotropic cosmologies,
its main physical results concern the description of the collapse of spherical configurations of matter. 
In this respect, one of the pioneering works was presented by Oppenheimer and Snyder \cite{PhysRev.56.455},
who considered a collapsing dust cloud with a homogeneous density, and concluded that all the matter eventually converges and forms a gravitational singularity. 
Indeed, this was a first step toward describing the formation of black holes. Subsequent studies, relaxing
symmetry restrictions and with more realistic matter, proved their intuition to be true: The gravitational singularity
is dynamically generated. The outcome of these investigations, synthesized in the singularity theorems \cite{PhysRevLett.14.57,hawking_ellis_1973,Senovilla:1998oua,Senovilla:2021pdg}, certifies the generally accepted incompleteness of general relativity (GR).
Providing a framework able to solve, or at least alleviate, this flaw
is in fact one of the main motivations for the search of a theory of quantum gravity.

In this context, the main goal of the present work is to
study an effective model that describes the spherical collapse of a dust cloud
with corrections motivated by loop quantum gravity.
Although the full quantum dynamics of this theory remains an open problem,
the homogeneous symmetry reduction of general relativity has been subject to a consistent loop quantization \cite{Ashtekar:2006wn,Ashtekar:2011ni,Ashtekar:2016ecx}. Further, the effective homogeneous
models implementing key aspects of the theory have been shown to greatly agree with the full quantum dynamics \cite{Taveras:2008ke,PhysRevD.87.043507,Agullo_2013} and, in particular, they predict
the resolution of the initial cosmological singularity.

The effective approach encodes the so-called holonomy corrections in the classical Hamiltonian constraint,
and thus modify the dynamics as predicted by general relativity.
However, the generalization of these studies to nonhomogeneous spacetimes has not been trivial.
In the canonical setting, the hypersurface deformation algebra encodes the covariance of the theory.
While
in homogeneous models this algebra is trivial because
the diffeomorphism constraint automatically vanishes, this is not the case in more general
scenarios, like under spherical symmetry. In such a case, a modification of the Hamiltonian constraint
leads in general to the nonclosure of the algebra, which signals a breaking of the covariance.

In the context of vacuum spherical symmetry, examples of deformed Hamiltonian constraints with a closed
hypersurface algebra were presented some time ago (see, for instance, \cite{Bojowald:2015zha,universe6030039,PhysRevD.102.106024,Alonso-Bardaji:2020rxb}). However, the corresponding structure function
does not transform adequately, and thus there is no known way to define an associated metric in
a covariant way (even if there were some proposals \cite{Bojowald:2018xxu},
the geometry they described turned out to be gauge dependent \cite{Alonso-Bardaji:2022ear}). Recently, 
in Refs.~\cite{AlonsoBardaji:2023bww,Bojowald:2023xat,Alonso-Bardaji:2023vtl}, a family of vacuum models
was presented ensuring, not only the closure of the algebra, but also the correct transformation
properties of the structure functions. These models are completely covariant and one
can construct the associated geometry, so that gauge transformations correspond to
coordinate transformations, though there is still some freedom in such construction, as will be detailed below.
A particular case of this family of vacuum models was already analyzed in detail in
Refs. \cite{Alonso-Bardaji:2021yls,Alonso-Bardaji:2022ear}, where it was
shown to provide a completely regular and geodesically complete spacetime. Such spacetime
can be understood as a regularized version of the Schwarzschild black hole, which, instead of the classical
singularity, presents a transition surface between a black-hole and white-hole region.
This model was also extended to charged spherical black holes in cosmological backgrounds \cite{Alonso-Bardaji:2023niu}.

For dynamical matter fields, the construction of covariant effective models with holonomy corrections has been more challenging \cite{LMMB14, Giesel:2021dug,PhysRevLett.128.121301,PhysRevD.106.024014,universe8100526,Giesel:2022rxi,Lewandowski:2022zce,Giesel:2023hys,Giesel:2023tsj}.
In Ref.~\cite{Alonso-Bardaji:2021tvy}, we presented a deformed Hamiltonian constraint for spherical gravity coupled
to a scalar matter field with a closed hypersurface deformation algebra. Furthermore,
in Ref.~\cite{Alonso-Bardaji:2023vtl} we proposed the minimal-coupling prescription to the vacuum metric as a systematic
construction of fully covariant models for any matter field, while in Ref.
\cite{Bojowald:2023djr} other proposals to couple matter that follow the same lines have been presented.

In this paper we will analyze the collapse of a spherical dust matter field, making use
of the model as given in Ref.~\cite{Alonso-Bardaji:2023vtl}.
We note that this has also been studied in~\cite{Duque:2023syb} with the conclusion of having a
singular geometry. Here, however, we will show that one can indeed covariantly associate a completely regular
family of geometries to the model. 

The rest of the paper is organized as follows. In Sec.~\ref{sec.phasespace} we present the model
in phase space, and solve the equations of motion by choosing the dust as the internal time
variable. In Sec.~\ref{sec.geometry} we study the geometry covariantly associated with the model
and use the freedom to consider a conformal factor in order to get a family of singularity-free
spacetimes. Sec.~\ref{sec.conclusions} summarizes the main results and conclusions of the study.
In App.~\ref{sec.appendix2} the family of metrics is presented in a different chart while in App.~\ref{sec.appendix1} we present an alternative model to describe a nonsingular dust collapse.

\section{The effective model}\label{sec.phasespace}

We have two pairs of conjugate variables for the geometric degrees of freedom, $\{\krad(x_1),\erad(x_2)\}=\delta(x_1,x_2)=\{\kangnew(x_1),\ephinew(x_2)\}$, and an additional one for the dust component, $\{\phi(x_1),P_\phi(x_2)\}=\delta(x_1,x_2)$. 
As commented above, the model we will consider
is constructed by minimally coupling the dust field
to the vacuum metric presented in Refs.~\cite{Alonso-Bardaji:2021yls,Alonso-Bardaji:2022ear}.
As explained in Ref.~\cite{Alonso-Bardaji:2023vtl},
this leads to the total Hamiltonian $ H_T=\int (N\ham  + N^x\diff) dx$, 
that describes the dynamics, where the two constraints
 \begin{subequations}\label{polham}
\begin{align}
\diff =&\, -\krad\erad' +\kangnew'\ephinew +\phi'\pphi,\label{eq.diff}
\end{align}
and
\begin{widetext}
\begin{align}\label{eq.ham}
\ham =&\, \frac{1}{\sqrt{1+\lambda^2}}\Bigg[ -\frac{{\ephinew}}{2\sqrt{{\erad}}}\left(1+\frac{\sin^2{{(\lambda {\kangnew})}}}{{{\lambda^2}}}\right)  -\sqrt{{\erad}}{\krad}\frac{\sin{(2\lambda {\kangnew})}}{\lambda}\left(1+\left(\frac{\lambda \erad'}{2{\ephinew}}\right)^{\!2}\right)\\
&+\left(\frac{(\erad')^2}{8\sqrt{{\erad}}{\ephinew}}  -\frac{\sqrt{{\erad}}}{2\ephinew^2}\erad'\ephinew'
 +\frac{\sqrt{{\erad}}}{2{\ephinew}}\erad'' \right)\cos^2{(\lambda {\kangnew})} \Bigg] +\pphi\sqrt{1+\frac{\cos^2(\lambda\kangnew)}{1+\lambda^2}\left(1+\left(\frac{\lambda \erad'}{2{\ephinew}}\right)^{\!2}\right)\frac{\erad}{\ephi^2}(\phi')^2} ,\nonumber
\end{align}
\end{widetext}
\end{subequations} 
satisfy the canonical hypersurface deformation algebra,
\begin{subequations}\label{eq.hda}
\begin{align}
    \{D[s_1],D[s_2]\}&=D[s_1s_2'-s_1's_2],\\
    \{D[s_1],H[s_2]\}&=H[s_1s_2'],\\
    \{H[s_1],H[s_2]\}&=D[F(s_1s_2'-s_1's_2)],\label{eq.hdahh}
\end{align}    
\end{subequations} 
with the structure function
\begin{align}\label{eq.structuref}
    F:=\frac{\cos^2(\lambda\kangnew)}{1+\lambda^2}\left(1+\left(\frac{\lambda \erad'}{2{\ephinew}}\right)^{\!2}\right)\frac{\erad}{\ephinew^2} ,
\end{align}
being non-negative since $E^x\geq 0$. 
The prime stands for the derivative with respect to $x$, and we have defined
the smeared form of the constraints $D[s]:= \int s\diff dx$ and $H[s]:=\int s\ham dx$. 
The real constant $\lambda\neq0$ is the polymerization parameter of the corrections motivated by loop quantum gravity, and GR is recovered in the limit $\lambda\to0$. 

In the following,
we will solve the six Hamiltonian equations of motion, $\dot{q}=\{q,H[N]+D[N^x]\}$ for $q=\krad,\kangnew,\phi,\erad,\ephinew,P_\phi$, with the dot being the time derivative,
along with the two constraint equations ${\cal D}=0$ and ${\cal H}=0$. 
As the first gauge choice, we impose
the dust field to be the time variable,
\begin{align}\label{eq.gaugetime}
    \phi={t},
\end{align}
and the conservation of this condition fixes the lapse
\begin{align}\label{eq.gaugelapse}
\lapse=1.
\end{align}
We then solve the diffeomorphism constraint equation, $\diff=0$, for $\krad=\kangnew'\ephinew/(\erad')$, which is valid provided $\erad'$ is nonidentically vanishing.
At this point it is convenient to define the functions
\begin{subequations}\label{eq.defs}
    \begin{align}
        r&:=\sqrt{\erad},\\
\label{eq.defm} \!\! m&:=\!\frac{\sqrt{\erad}}{2}\!\left(1+\!\frac{\sin^2(\lambda\kangnew)}{\lambda^2}-\!\left(\frac{\erad'}{2\ephinew}\right)^{\!2}\!\!\cos^2(\lambda\kangnew)\!\right)\!\!,\\
           {\kappa}&:=\left(\frac{\erad'}{2\ephinew}\right)^2-1,
    \end{align}
\end{subequations}
which will greatly simplify the equations. Note that these are just definitions,
which imply $\kappa\geq -1$ and $r\geq 0$, but
no gauge fixing is involved until we choose the specific form of some of these functions. 

Since $\phi$ is the time variable, we solve $\ham=0$ for its conjugate variable
$P_\phi$, which provides, in this gauge, a notion of conserved energy
\begin{align}\label{eq.pphi}
    \pphi={\sqrt{1-\lambdabar}}\frac{m'}{\sqrt{1+\kappa}}\,,
\end{align}
with $\lambdabar:=\lambda^2/(1+\lambda^2)\in(0,1)$ being a bounded constant that,
as it will be shown below,
measures the strength of the quantum corrections at the geometric level.
Note that the limit $\lambdabar\to0$ stands for GR.
Furthermore, in terms of these variables, the structure function \eqref{eq.structuref}
takes the simple form,
\begin{equation}\label{eq.Fsolution}
 F=\frac{(1+\kappa)}{(r')^2}\left(1-\frac{2\lambdabar m}{r}\right),
\end{equation}
which, as explained in Ref.~\cite{Alonso-Bardaji:2022ear}, is directly related to the inverse of the radial component of the metric. In fact, the most general line element covariantly associated with the Hamiltonian \eqref{polham} is given below in Eq. \eqref{eq.metricgen}, with $N=1$ for this particular gauge. Since, as commented above, $F$ cannot take negative values,
the dynamical functions are restricted by $\kappa\geq -1$, which was implicit
in the change of variables above, and by $r\geq 2\lambdabar m$.
The saturation of these inequalities corresponds to the vanishing of $F$.
More precisely, the root $(1+\kappa)/(r'^2)=0$ corresponds to $E^x/(E^\varphi)^2=0$,
while $r= 2\lambdabar m$ implies $\cos(\lambda K_\varphi)=0$. Interestingly,
the latter does not exist in GR, and it corresponds to a symmetry-reflection
point of the Hamiltonian constraint.

The remaining three equations of motion read
\begin{subequations}\label{eq.dots}
\begin{align}
 \dot{r}&=\shift r'-\epsilon\sqrt{1-\frac{2\lambdabar m}{r}}\sqrt{{\kappa}+\frac{2m}{r}},\label{eq.dotr}\\
 \label{eq.dotm}  {\dot{m}}&=\shift{m'},\\[4pt]
    \label{eq.dotk} \dot{{\kappa}}&=\shift {\kappa}',
\end{align}
\end{subequations}
where $\epsilon:=-\mathrm{sgn}\big(\lambda\sin(2\lambda\kangnew)\big)$
and we have assumed, without loss of generality, $\lambda K_\varphi\in[0,\pi)$. 
Note that \eqref{eq.dotr}
implies $r\geq 2\lambdabar m$ and $\kappa\geq -2 m/r$.
The saturation of these inequalities $r= 2\lambdabar m$ and $r= -2 m/\kappa$
corresponds to the points $\cos(\lambda\kangnew)=0$ and $\sin(\lambda\kangnew)=0$,
respectively, where $\epsilon$ changes sign.

Since the diffeomorphism
constraint has already been solved for $K_x$, 
\eqref{eq.dots} is a system of three equations for four variables
$N^x$, $r$, $m$, and ${\kappa}$. 
Therefore, we still have the gauge freedom to choose one function.
More precisely, the gauge would be completely fixed by choosing the specific form
of one among the functions $r$, $m$, ${\kappa}$
(or a combination between them), in such a way that its conservation provides the
shift $N^x$. 
In the following, we will solve the equations of motion in a particular gauge,
which will allow us to obtain an analytic solution for $r(t,x)$.

Let us require the mass function to be independent of the time, $m=m(x)$, with $m'(x)$ not being
exactly vanishing (because a constant $m$ restricts us to the vacuum case).
From \eqref{eq.dotm} one then gets $\shift=0$, which also imposes $\kappa=\kappa(x)$
as a solution to \eqref{eq.dotk}. In this way,
the only remaining equation reads
\begin{align}\label{eq.dotra}
    \dot{r}=-\epsilon\sqrt{1-\frac{2\lambdabar m}{r}}\sqrt{{\kappa}+\frac{2m}{r}},
\end{align} 
with two arbitrary functions $m=m(x)$ and $\kappa=\kappa(x)$, which parametrize each solution. 
In this gauge, the sign $\epsilon$ completely
determines the sign of $\dot{r}$ and, thus, whether we are in a collapsing ($\epsilon=1$) or expanding ($\epsilon=-1$) scenario.
In addition, $r= 2\lambdabar m$ and $r=-2 m/\kappa $ will define turning points of the trajectory
$r=r(t,x)$ as a function of $t$,
as long as the right-hand side of these expressions is positive because, as commented above,
$r\geq0$. 

More precisely,
in this gauge the coordinate $x$ labels the different layers of the dust and, for physically realistic scenarios, we expect $m(x)$ to be strictly positive (indeed, monotonically increasing) except
at the origin, that is, we set $m(0)=0$ and $m(x\neq0)>0$. 
Taking into account the sign of the second derivative, $r_{\rm min}(x):=2\lambdabar m(x)$ will be a minimum of the radius
$r$ for the layer $x\neq0$,
\begin{align*}
\ddot{r}\big{|}_{r=2\lambdabar m} &=\frac{(1+\lambdabar \kappa)}{4 m\lambdabar^2}>0,
\end{align*}
while  $r_{\rm max}(x):=-2 m/\kappa $ will define
a maximum,
\begin{align*}
\ddot{r}\big{|}_{r=-2 m/\kappa} &=-\frac{\kappa^2}{4m}(1+\lambdabar \kappa)<0,
\end{align*}
provided $\kappa(x)$ is strictly negative there.

As a result,
if they are initially collapsing $(\epsilon=-1)$, all the layers with $x\neq 0$
will bounce at $r=r_{\rm min}(x)$. In this scenario, the maximum $r_{\rm max}(x)$ will be finite
only if $\kappa(x)$ is negative, which defines the bounded case.
If $\kappa(x)$ is non-negative (the so-called unbounded case for $\kappa>0$
and marginally bounded for $\kappa=0$), the layer $x$ will reach
$r\to \infty$ with velocity $\sqrt{\kappa(x)}$.
For $x=0$, where $m(0)=0$,
one will obtain a bounce at $r=0$, though the function $r$ will in general not be differentiable there.
Let us see this in more detail by solving the evolution equation \eqref{eq.dotra}.

Taking into account that only a derivative with respect to $t$ appears in \eqref{eq.dotra}, and that it is separable, it is straightforward to reduce
the equation to an integral. This can be explicitly computed to obtain the following
implicit solution for $r=r(t,x)$, with the integration function $t_0=t_0(x)$:
\begin{subequations}\label{eq.t(r)}
\begin{align}\label{eq.t(r)k>0}
   t-t_0 =&\,  -\epsilon\frac{r}{\kappa}\sqrt{1-\frac{2\lambdabar m}{r}}\sqrt{\frac{2m}{r}+{\kappa}} +\epsilon\frac{2m}{{\kappa}^{3/2}}(1-\lambdabar\kappa)\mathrm{artanh}\sqrt{\frac{\kappa(r-2\lambdabar m)}{2m+\kappa r}},
\end{align}
for the unbounded case $\kappa>0$,
\begin{align}\label{eq.t(r)k=0}
   t-t_0 =-\epsilon\sqrt{\frac{{2} r^{3}}{9{m}}}\sqrt{1-\frac{2\lambdabar m}{r}}\left(1+\frac{4\lambdabar m}{r}\right),
\end{align}
for the marginally bounded case $\kappa=0$, and
\begin{align}\label{eq.t(r)k<0}
   t -t_0=&\,+ \epsilon\frac{r}{|\kappa|}\sqrt{1-\frac{2\lambdabar m}{r}}\sqrt{\frac{2m}{r}-|\kappa|}  - \epsilon\frac{2m}{|\kappa|^{3/2}}(1+\lambdabar|\kappa|)\arctan\sqrt{\frac{|\kappa|(r-2\lambdabar m)}{2m-|\kappa| r}},
\end{align}
for the bounded case $\kappa<0$.
\end{subequations}

As can be checked explicitly from above,
if $m(x)>0$, $r(t,x)$ is an everywhere analytic function of $t$, and has a minimum at $t=t_0$. 
For the shell $x=0$ with $m(x)=0$, we find that $\kappa(x)$ cannot be negative, and $r=\sqrt{\kappa}|t-t_0|$,
which is continuous for all $t$ and has also a minimum at $t=t_0$.
However, it is not differentiable there, unless we also fix the boundary condition $\kappa(0)=0$. This condition means that the ``central'' shell $x=0$ remains still at a constant radius.

Once we choose an initial value for $\epsilon$, let us say, for instance,
$\epsilon=1$ at a given time $t$ so that we are in the contracting branch,
the above solution is valid for $t\leq t_0$.
At $t=t_0$, which corresponds to $ r=r_{min}(x)=2\lambdabar m(x)$, we reach the minimum of the trajectory.
From that point on, we enter the expanding branch, where $\epsilon=-1$.
One can indeed check this by computing $\dot{\kangnew}$ at $\cos(\lambda\kangnew)=0$, to see that it is not zero there, and thus
$\epsilon=-\mathrm{sgn}\big(\lambda\sin(2\lambda\kangnew)\big)$ must change sign at $t=t_0$.

Only in the bounded case, with $\kappa<0$,
there is a finite maximum $r_{\rm max}(x)=2m/|\kappa|$ for the trajectories $r(t,x)$. This means that the shell $x$ follows a periodic dynamics, oscillating
between its corresponding $r_{\rm min}$ and $r_{\rm max}$, with period $T=2\pi (1+\lambdabar|\kappa|)m/|\kappa|^{3/2}$. The sign of $\epsilon$ changes at both $r_{min}$ and $r_{max}$, as they correspond to $\cos(\lambda\kangnew)=0$ and $\sin(\lambda\kangnew)=0$, respectively.

In summary, the solution \eqref{eq.t(r)} describes a function $r=r(t,x)$ with a local
minimum at $r=2\lambdabar m$ for $m>0$. In such a case,
$r$ is everywhere analytic as a function of $t$. 
This extends to $m(0)=0$, which corresponds to the value of the mass function at the origin, provided the boundary condition $\kappa(0)=0$ is enforced.

In the following section, we will present the geometric description of the system and
show that one can endow the model with a regular
geometry.

\section{The geometric description}\label{sec.geometry}

In spherical symmetry, the spacetime manifold ${\cal M}$ is a warped product between a two-dimensional Lorentzian manifold ${\cal M}^2$ and the two-sphere $S^2$. The function $x$, constant on the orbits of the spherical symmetry group, defines a radial direction outside the fixed points of the group. 
Choosing adapted coordinates, the spacetime metric is diagonal by blocks,
\begin{align}\label{eq.metricgen1}
    ds^2=g_{AB}dy^A dy^B+R^2 d\sigma^2,
\end{align}    
with $d\sigma^2$ the metric of the unit two-sphere. 
We will use $t:=y^0$ and $x:=y^1$. The area-radius function $R=R(t,x)$ encodes the area of the two-spheres, and 
we assume that $\mathcal{M}^2$ is foliated by the spacelike level surfaces of the time function $t$. 

\subsection{The construction of the metric}

To endow the model with a metric of the form \eqref{eq.metricgen1}, one needs to ensure that gauge transformations on phase space correspond to coordinate changes in the spacetime manifold.
As explained in Ref.~\cite{Alonso-Bardaji:2022ear}, the metric
\begin{align}\label{eq.metricgen2}
    ds^2=-{N}^2dt^2+\frac{1}{F}\big(dx+N^x dt\big)^2+r^2 d\sigma^2,
\end{align}
where $F$ is the structure constant \eqref{eq.structuref} fulfills such requirement, and provides a covariant geometric representation
of the model. The main reason is that, under the gauge transformations generated by the first-class constraints~\eqref{polham}, $1/F$ transforms in the same way as a spatial metric under an infinitesimal
coordinate transformation. 
That is, $F$, as given in \eqref{eq.structuref}, obeys the relation
\begin{align}\label{eq.covariance}
&\xi^t\partial_t\left({\frac{1}{F}}\right)+\xi^x\partial_x \left(\frac{1}{F}\right) +\frac{2}{F}\left({N^x}\partial_x{\xi^t}+\partial_x{\xi^x}\right)
=\left\{\frac{1}{F},H\big[ \xi^t N\big]+D\big[\xi^tN^x+\xi^x\big]\right\}
\end{align}
for any parameters $\xi^t$ and $\xi^x$, when the equations of motion are satisfied.

In principle, if we had the complete
form of the hypersurface deformation algebra (without any symmetry reduction), one could read
the whole inverse spatial metric $q^{ij}$ from the bracket \eqref{eq.hdahh}. 
However,  
the vanishing of the angular components of the diffeomorphism constraint in spherical symmetry 
prevents us from determining the metric in $S^2$.
Therefore, to write down \eqref{eq.metricgen2},
we have assumed that the area of the spheres of the spherical symmetry
is the same as in GR and thus we have chosen $R:=r=\sqrt{\erad}$.
Nonetheless,
one could choose any other scalar field multiplying the $S^2$ metric.

As we will show below, the metric \eqref{eq.metricgen2} turns out to have a diverging
curvature at the turning point $r_{min}=2\lambdabar m$ for the dynamics described in the previous section. 
Therefore, at this point, we recall another freedom in the construction
of the geometry associated with the model, which will allow us to resolve such
divergences. Indeed,
as explained in Refs.~\cite{Bojowald:2018xxu,Alonso-Bardaji:2022ear} 
the sector of the metric corresponding to ${\cal M}^2$ can be multiplied by 
a conformal factor,
\begin{align}\label{eq.metricgen}
    d\tilde{s}^2={\Omega^{-2}}\left[-{N}^2dt^2+\frac{1}{F}\big(dx+N^x dt\big)^2\right]+r^2 d\sigma^2,
\end{align}
and still obey the covariance conditions as long as $\Omega$ is a spacetime scalar field.
The main reason is that multiplying the different objects by a scalar does not
change their transformation properties. In particular, if $F$ obeys relation
\eqref{eq.covariance}, so will $\Omega^2 F$ for any scalar $\Omega$.
Note that no structure on the phase space (neither the constraints,
nor the equations of motion) depend on the scalar $\Omega$. This is a freedom that appears when endowing the model with a geometric description, which is also present in canonical GR. { That is, given the GR hypersurface
deformation algebra in spherical symmetry, from there one would be able to infer the metric in the two-dimensional
Lorentzian sector, only up to a conformal factor. The freedom to choose this conformal factor is not usually considered, and
it is completely fixed if one demands that the metric satisfies the Einstein equations.}
However, since, contrary to GR, we lack the fundamental field equations, we have no clear indication to choose one among the infinite family of conformal metrics. We thus rely on singularity resolution to find the suitable conformal factor for the Lorentzian $\mathcal{M}^2$ sector.

\subsection{Curvature scalars}

In such spherically symmetric spacetimes \eqref{eq.metricgen1}, all the information regarding the spacetime curvature is encoded in the norm $v_Av^A$ of the vector $v^A:=\nabla^A R$, the trace of its gradient $\nabla_A v^A$, and the Ricci scalar ${}^{(2)}\mathcal{R}$ of the Lorentzian metric $g$ in $\mathcal{M}^2$. More precisely, there are only two independent spacetime curvature scalars: 
\begin{subequations}\label{eq.scalars}
    \begin{align}
\!\!{}^{(4)}\mathcal{R}&={}^{(2)}\mathcal{R}+\frac{2}{R^2}\big(1-v_Av^A\big)-\frac{4}{R}\nabla_Av^A,\\
\!\!{\cal U}&=-\frac{1}{6}\left({}^{(2)}\mathcal{R}+\frac{2}{R^2}\big(1-v_Av^A\big)+\frac{2}{R}\nabla_Av^A\right),
\end{align}
\end{subequations}
with the former being the four-dimensional Ricci scalar, and the latter providing all nonvanishing components of the Weyl tensor. 

Now, given two conformal metrics 
$\widetilde{g}=\Omega^{-2}g$ in $\mathcal{M}^2$, we have that 
$\widetilde{v}_A\widetilde{v}^A=\Omega^2v_Av^A$, $\widetilde{\nabla}_A\widetilde{v}^A=\Omega^2\nabla_Av^A$, and ${}^{(2)}\widetilde{\mathcal{R}}=\Omega^{2}\big({}^{(2)}\mathcal{R}+\nabla^A\nabla_A\log\Omega\big)$. 
Therefore, we see that a sufficiently rapidly vanishing $\Omega$ flattens the metric in $\mathcal{M}^2$. When translated to the four-dimensional spacetime,
\begin{subequations}
    \begin{align}
\!\!{}^{(4)}\widetilde{\mathcal{R}}&=\Omega^2\left({}^{(4)}\mathcal{R}+\nabla^A\nabla_A\log\Omega\right)+\frac{2}{R^2}\big(1-\Omega^2\big),\\
\!\!\widetilde{\cal U}&=\Omega^2\left(\mathcal{U}-\frac{1}{6}\nabla^A\nabla_A\log\Omega\right)-\frac{1}{3R^2}\big(1-\Omega^2\big),
\end{align}
\end{subequations}
and we can exploit this conformal freedom to remove curvature divergences 
without changing the dynamics on phase space. 

In fact, as already commented above, using \eqref{eq.metricgen2} along with the solutions \eqref{eq.gaugelapse}, \eqref{eq.Fsolution}, and \eqref{eq.dots}, one can check that the curvature is divergent at the surfaces $r=2\lambdabar m$.
More precisely, as $r\to2\lambdabar m$, we find $v_Av^A\sim\beta$, $\nabla_Av^A\sim\beta^{1/2}$, and ${}^{(2)}\mathcal{R}\sim\beta^{-1}$, where the scalar
\begin{align}\label{eq.beta}
    \beta=1-\frac{2\lambdabar m}{r}
\end{align}
is the ``deformation'' of the structure function with respect to GR. That is, $\beta:=F/(F|_{\lambda\to0})$. This result might come as a bit of a surprise since, as explained above,
in the phase space all trajectories present a smooth and continuous
bounce at $r=2\lambdabar m$.
However, choosing a different conformal factor, let us say $\Omega^2=\beta^n$, with $n\geq1$, is enough to get finite spacetime curvature scalars at the hypersurfaces $r=2\lambdabar m$.
 Note that, since $\beta$ trivially reduces to one as $\lambdabar\to0$,
 with this choice 
we still recover GR in that limit.
In addition, we would like to point out that the case $n=1$ is special because it is the only one in which
${}^{(2)}\widetilde{\mathcal{R}}$ does not vanish there.

\subsection{The effective geometry}

In the following, we will use the metric \eqref{eq.metricgen}, with the conformal factor $\Omega^2=\beta^n$, and the solutions for the phase-space variables \eqref{eq.gaugelapse}, \eqref{eq.Fsolution}, and \eqref{eq.dots}
found in Sec.~\ref{sec.phasespace} for a particular gauge. In this gauge, the metric \eqref{eq.metricgen} reads
\footnote{As a particular example of the covariance of the model,
in App.~\ref{sec.appendix2}
we provide the solution of the equations of motion in a different gauge,
construct the corresponding line element, and show that it is related to this line element
by a specific coordinate transformation.
}
\begin{align}\label{eq.metricdiag}
    d\tilde{s}^{2}=&\,-\left(1-\frac{2\lambdabar m(x)}{r(t,x)}\right)^{-n}{dt}^{2} + \left(1-\frac{2\lambdabar m(x)}{r(t,x)}\right)^{-(n+1)}\frac{\big(r'(t,x)\big)^{2}}{1+{\kappa(x)}}{dx}^{2} +r(t,x)^{2}{d\sigma}^{2},
\end{align}
with $n\geq1$, $m=m(x)$, $\kappa=\kappa(x)$,  
and $r(t,x)$ implicitly given by \eqref{eq.t(r)}. 
This describes a regular spacetime for $x>0$. { In particular, at the minimum of the trajectories $r=2\lambdabar m$, the Ricci scalar is ${}^{(4)}\mathcal{R}=(1+\lambdabar\kappa)/(2\lambdabar^3m^2)\delta_n^1+2/(2\lambdabar m)^2$.} However, the situation of the layer $x=0$, where $m(x)=0$, is a bit different. If the function $r(t,0)$ is positive there, the curvature vanishes. If $r(t,0)=0$, the leading contribution as $r\to0$ to the Ricci scalar ${}^{(4)}{\mathcal{R}}$ goes as $m/r^3$, which does not generically provide a convergent curvature. Indeed, we need that $m$ decays at least as $r^3$. More precisely, if we have $r\approx a(t)x$ and $m\approx \rho_0 x^3$ in a neighborhood of $x=0$, we find that ${}^{(4)}\mathcal{R}=6\rho_0\big(1+\lambdabar(2+\kappa)\big)/a(t)^3$ at $x=0$.

We can physically motivate such decay by introducing the Hawking mass $M_H$,
which measures the amount of energy contained within a sphere of constant coordinates $t$ and $x$. This notion is tightly related to the norm of $v_A$ (see below), and it reads $M_H:=(1-v_Av^A)r/2$, which leads to
\begin{align}\label{eq.Hmass}
     M_H=\frac{r}{2}\left[1-\left(1-\frac{2m}{r}\right)\left(1-\frac{2\lambdabar m}{r}\right)^{n+1}\right].
\end{align}
Since $M_H=\lambdabar m$ at $r=2\lambdabar m$, the function $m$ inherits the physical meaning of mass at those hypersurfaces. Therefore, the homogeneous-density case $M_H=\frac{1}{6}\rho(t)r^3$ implies the above decaying conditions. In such a case, the geometry is everywhere regular.

To close the section, let us specify the { dust} energy density,
defined as \eqref{eq.pphi} divided by the determinant of the spatial metric, i.e.,
\begin{align}\label{eq.densitydust}
    \mathcal{E}:=\frac{\sqrt{\Omega^2F}}{r^2}\pphi =\sqrt{1-\lambdabar}\frac{ m'}{r^2r'}\left(1-\frac{2\lambdabar m}{r}\right)^{\frac{n+1}{2}} ,
\end{align}
which vanishes at $r=2\lambdabar m$ because, as one can read from \eqref{eq.defm}, $r'=2\lambdabar m'$ when $r=2\lambdabar m$. 

{ Outside the surface of the star, i.e., when $m'=0$ so that \eqref{eq.densitydust} vanishes, we see that the Ricci scalar ${}^{(4)}\mathcal{R}$ decreases as $1/r^4$ when $r\to\infty$ for any $n\geq1$.}

\subsection{The structure of the spacetime}

\begin{figure}[t]
    \centering
        \includegraphics[width=0.75\textwidth]{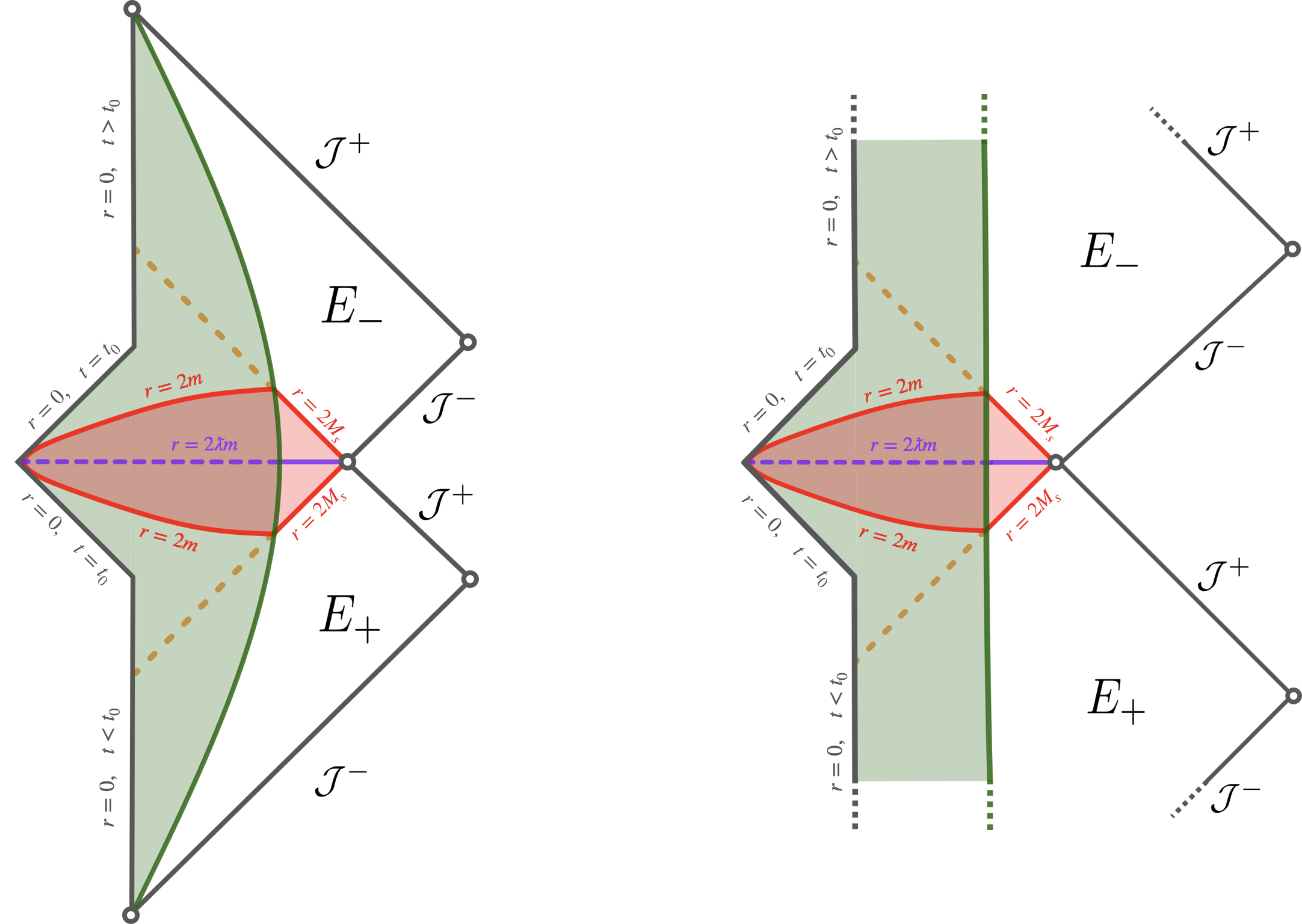}
     \caption{Outline of the conformal diagram of the spacetimes under consideration (the metric \eqref{eq.metricdiag} with positive $m(x)$ for all $x>0$). On the left, $\kappa\geq0$ and $\kappa<0$ on the right. In the latter, the star oscillates periodically between a minimum and a maximum radius. In both cases, the green line represents the surface of the star (drawn by hand, in particular in the bounded case, where we depict it as a vertical line), and the region on its right describes a singularity-free vacuum solution similar to  the geometry analyzed in \cite{Alonso-Bardaji:2022ear}. 
    The red lines correspond to the apparent
    horizons $r=2m$, with their interior shaded in red. The purple horizontal line is the regular minimal spacelike transition surface $r=2\lambdabar m$, which divides this interior into a trapped (below) and antitrapped (above) regions. The dashed yellow line is the
     null cone that defines the event horizon.}
    \label{fig:diagram}
\end{figure}

Note that, even if from a dynamical perspective our gauge was perfectly valid at $r=2\lambdabar m$, the metric in this chart, as given by \eqref{eq.metricdiag}, degenerates at those surfaces. This is however just a coordinate singularity,
and a chart covering them should exist. We have been unable to explicitly construct
such a chart, though we can {{deduce}} the structure of the spacetime by constructing scalar quantities as follows.

Since $v_A dx^A=\dot{r}dt+r'dx$, all the components of its metrically conjugate vector, $v^A=g^{AB}v_B$, vanish at $r=2\lambdabar m$. To be more precise, we can project it along the unit normal to the leaves of constant time, $n_A dx^A$,
and its orthogonal direction, 
\begin{subequations}
\begin{align}
 \label{eq.nava}   n_A v^A = -\epsilon \sqrt{\kappa+\frac{2m}{r}}\left(1-\frac{2\lambdabar m}{r}\right)^{\frac{n+1}{2}} ,\\
    \epsilon_{AB } n^A v^B = -\sqrt{1+\kappa}\left(1-\frac{2\lambdabar m}{r}\right)^{\frac{n+1}{2}}.
\end{align}
\end{subequations}
It is clear that these components are vanishing at  $r=2\lambdabar m$,
which shows that the hypersurfaces defined as $r=2\lambdabar m$ are minimal. Besides, if $n>1$, ${}^{(2)}\mathcal{R}$
is also zero there, and thus they are flat.

In addition, these surfaces are spacelike, since in a neighborhood of $r=2\lambdabar m$, the norm of its gradient,
\begin{align*}
 \nabla^A(r-2\lambdabar m)\nabla_A(r-2\lambdabar m)=\left(1-\frac{2\lambdabar m}{r}\right)^{n+1}
 \left[\left(1-\frac{2 m}{r}\right)+\frac{4\lambdabar m'}{(r')^2}(\lambdabar m'-r')(1+\kappa) \right],
\end{align*}
is negative, because, from \eqref{eq.defm}, we find $r'=2\lambdabar m'$ when $r=2\lambdabar m$.
The norm of the vector $v^A$,
\begin{align}\label{eq.normv}
    v_A v^A =\left(1-\frac{2 m}{r}\right)\left(1-\frac{2\lambdabar m}{r}\right)^{n+1} ,
\end{align}
vanishes at $r=2 m$ in addition to $r=2\lambdabar m$. 
The former defines the apparent horizon,
where the vector becomes lightlike,
while the latter corresponds to the commented minimal surfaces, where the vector vanishes
and the bounce of the dust shells takes place.
Going back to \eqref{eq.nava}, we see that $v^A$ points to the future (past) when $\epsilon=1$ ($\epsilon=-1$), that is, in the contracting (expanding) phase. Since $v_Av^A<0$ for $2\lambdabar m<r<2m$, this region is trapped to the future (past), whereas the regions $r>2m$ are nontrapped.
Note, in particular, that, since $\lambdabar<1$,
the bounce always takes place in the trapped region
inside the horizon and there is no mass threshold to form the horizon.

In addition, the surface $r=2\lambdabar m$ is a symmetry surface of the
complete spacetime, because it corresponds to the symmetry surface $\cos(\lambda\kangnew)=0$ of the Hamiltonian \eqref{polham}. Taking all this information into account, in Fig.~\ref{fig:diagram}
we present the qualitative conformal diagram for the family of metrics \eqref{eq.metricdiag}.
This diagram can be obtained from its classical counterpart, just by replacing
the singularity surface by the now smooth minimal spacelike surface $r=2\lambdabar m$, 
and extending the spacetime further away than this surface by reflection symmetry.

\section{Concluding remarks}\label{sec.conclusions}

We have analyzed a model that describes the collapse of a spherical dust
cloud with corrections motivated by loop quantum gravity.
This model is constructed by introducing certain corrections in
the Hamiltonian constraint of general relativity, while requesting the closure of the hypersurface 
deformation algebra and a proper transformation of the structure functions under gauge changes.
This allows us to preserve the covariance of the theory
and to construct an associated geometry in a gauge-invariant way, i.e.,
so that each gauge choice in the phase space simply corresponds to a
choice of coordinates in spacetime. 

Fixing the dust as the internal time variable, we have shown that,
given a positive mass function,
the solution to the equations of motion generically leads to a minimum value
of the radius for each dust shell.
Contrary to the dynamics described by Einstein equations, where
initially collapsing shells end up reaching $r=0$ and producing a
singularity, in this model all the shells bounce back at their corresponding
positive minimum.

Then, we have proceeded to analyze the associated geometry of the model.
However, it turns out that the metric \eqref{eq.metricgen2} presents a curvature divergence
at the bouncing point, even if dynamical trajectories in phase space
are completely smooth and continuous there. This metric was the one considered
in previous analysis of vacuum models \cite{Alonso-Bardaji:2021yls,Alonso-Bardaji:2022ear,Alonso-Bardaji:2023niu},
and was indeed used in Ref. \cite{Alonso-Bardaji:2023vtl} to construct the matter Hamiltonian of the present model
by following a minimal-coupling prescription.
Nonetheless, this is not the only geometry one can associate with the effective
model, and there is, in particular, the freedom to introduce a conformal
factor, as long as it is a spacetime scalar.

Therefore, motivated by the fact that the dynamics is completely smooth in
phase space, we have introduced the metric \eqref{eq.metricgen}, and
found out the necessary conditions for the scale factor $\Omega$, so that it defines
an everywhere smooth geometry. In this way, we have ended up with the family
of metrics \eqref{eq.metricdiag}, with $n\geq 1$.
Given a mass function positive everywhere except at the origin, where it vanishes
but scales as $r^3$, these metrics
describe smooth geometries covariantly associated with the model.
In Fig. \ref{fig:diagram} we have qualitatively constructed their conformal
diagram.

In summary, one would consider \eqref{eq.metricdiag} with $n\geq 1$
as the physical metric, while the dust field is minimally coupled to
the metric \eqref{eq.metricgen2}, which corresponds to $n=0$ in \eqref{eq.metricdiag}.
In principle, minimal coupling does not need to be
more fundamental than any other coupling. However, in App. \ref{sec.appendix1},
we also show that it is indeed possible to obtain a similar model, where
the dust field is minimally coupled to the physical smooth metric. 

The main difference between the present model and other effective descriptions
of the dust collapse in the context of loop quantum gravity (see, for instance, Refs.~\cite{PhysRevD.106.024014,Giesel:2022rxi}, where scale-dependent holonomies ($\bar{\mu}$-scheme) are considered) lies in the emphasis
of the covariance of the theory by ensuring that the polymerized constraint \eqref{polham} satisfies
the hypersurface deformation algebra \eqref{eq.hda}. While this is a key feature to consistently explore the modified geometries, the resulting Hamiltonian \eqref{polham} is more complicated than what one would expect from the usual holonomy corrections, even if our polymerization parameter is constant ($\mu_0$-scheme). Concerning the physical implications, the typical curvature
corrections to the GR potential outside the star differ and,
while the asymptotic decay of the Ricci scalar goes as $1/r^6$
in the models analyzed in Refs.~\cite{PhysRevD.106.024014,Giesel:2022rxi},
in the present framework it scales as $1/r^4$. Furthermore, in our model
the bounce of the dust shells always happens
in the trapped region inside the apparent horizon, and thus there is neither a mass threshold
to form such horizon as in Ref.~\cite{Giesel:2022rxi}, nor an interior Cauchy horizon before the bounce
as in Ref.~\cite{PhysRevD.106.024014}.
These key features are robust and they do not rely on the specific initial density profile.

Finally, we note that general relativity is recovered
in the limit where the parameter $\lambdabar$, which measures the strength of
the quantum-gravity corrections, tends to zero (or, equivalently, $\lambda\to0$).
All the presented metrics, along with the equations of motion \eqref{eq.dots},
reproduce the classical LTB results in that limit, including the gravitational singularity as $r\to0$.
In this sense, general relativity can be considered as
a singular limit of the model.

\section*{Acknowledgments}

This work was supported by the Basque Government Grant IT1628-22, and by the Grant PID2021-123226NB-I00 (funded by MCIN/AEI/10.13039/501100011033 and by “ERDF A way of making Europe”). A.A.B.'s work was made possible through the support of the ID\# 62312 grant from the John Templeton Foundation, as part of the project \href{https://www.templeton.org/grant/the-quantum-information-structure-of-spacetime-qiss-second-phase}{``The Quantum Information Structure of Spacetime'' (QISS)}. The opinions expressed in this work are those of the authors and do not necessarily reflect the views of the John Templeton Foundation.

\appendix

\section{An alternative gauge/chart}\label{sec.appendix2}

In this appendix we provide the metric \eqref{eq.metricdiag} in an alternative chart in order
to explicitly illustrate the covariance of the model. We will solve the equations of motion for
another gauge, construct the corresponding line element, and then show that they can be related to the
line element \eqref{eq.metricdiag} by a coordinate transformation.

We now set $r=x$ in \eqref{eq.dots}. Then, $N^x=\epsilon\sqrt{1-2\lambdabar m/r}\sqrt{\kappa+2m/r}$, which determines the evolution of $m(t,x)$ and $\kappa(t,x)$ through \eqref{eq.dotm} and \eqref{eq.dotk}, respectively.
Plugging this into the metric \eqref{eq.metricgen} leads to the line element
\begin{align}\label{eq.metricGP}
    &d\tilde{s}^{2}=\frac{1}{1+{\kappa}}\left(1-\frac{2\lambdabar m}{r}\right)^{-n}\Bigg[-\left(1-\frac{2m}{r}\right){dt}^{2}+2\epsilon\sqrt{\frac{{\kappa}r+{2m}}{r-2\lambdabar m}}dtdr
   +\left(1-\frac{2\lambdabar m}{r}\right)^{-1}{dr}^{2} \Bigg]+r^{2}{d{\sigma}}^{2}. 
\end{align}
Besides, it is straightforward to see that
the change of coordinates $dr=r'dx+\dot{r}dt$, with $\dot{r}$ given by \eqref{eq.dotra}, renders \eqref{eq.metricdiag} into \eqref{eq.metricGP}. Therefore, this shows that these two different gauge choices lead to different charts of the same metric. 

Furthermore, this chart provides an interesting characterization of the function $m(t,x)$, since the time-time component of the Einstein tensor, which can be interpreted as certain energy density, yields
\begin{align}
    \rho:=-G_0{}^0 =\frac{2}{r^2}\frac{\partial M_H}{\partial r},
\end{align}
with $M_H$ the Hawking mass \eqref{eq.Hmass}.

\section{Minimal coupling of the dust field to the physical metric}\label{sec.appendix1}

As commented in the main text, we can understand the above dust matter model
as being minimally coupled to a fiducial conformal metric. That is,
the physical smooth metric is \eqref{eq.metricdiag} with $n\geq 1$,
while the dust is minimally coupled to the metric
\eqref{eq.metricgen2}, which corresponds to $n=0$ in \eqref{eq.metricdiag}.
However, in this appendix we will show that it is indeed possible to construct an alternative model
such that the dust is minimally coupled to the physical smooth metric.

Let us define the vacuum Hamiltonian as
\begin{align}
    H_T^{(grav)} = \int \big(N^x\diff_g+N\ham_g)dx,
\end{align}
with $\diff_g$ and $\ham_g$ being the reduction to vacuum of \eqref{eq.diff} and \eqref{eq.ham}, respectively,
that is, the ones obtained by setting $P_\phi=0$ in those expressions.
Now, minimally coupling the dust to the metric \eqref{eq.metricgen} gives the matter contribution
\begin{align}
    H_T^{(dust)} =\int\Big(N^x\phi'P_\phi+ N\Omega^{-1}P_\phi\sqrt{1+F\Omega^2 (\phi')^2}\Big)dx,
\end{align}
to the total Hamiltonian $H_T=H_T^{(grav)}+H_T^{(dust)}$. 
Note how the conformal factor $\Omega$ enters the dust Hamiltonian constraint, meaning that the equations
of motion of the different variables will change with respect to the model studied in the main text.
In order to remain as close as possible to our previous results, we will set $\Omega^2=\beta^n$, with $\beta$ given in \eqref{eq.beta}, in the following.

Using again the definitions \eqref{eq.defs} and setting the dust as time, $\phi={t}$, leads now to
the lapse $\lapse=\Omega$, and to the following set of equations,
\begin{subequations}\label{eq.dots2}
\begin{align}
 \dot{r}&=\shift r'-\epsilon\left({1-\frac{2\lambdabar m}{r}}\right)^{(n+1)/2}\sqrt{{\kappa}+\frac{2m}{r}},\label{eq.dotr2}\\
 \label{eq.dotm2}  {\dot{m}}&=\shift{m'},\\
    \label{eq.dotk2} \dot{{\kappa}}&=\shift {\kappa}'
    -{\epsilon}\lambdabar{n}(1+\kappa)\frac{{2}m}{r^2}\left({1-\frac{2\lambdabar m}{r}}\right)^{(n-1)/2}\sqrt{{\kappa}+\frac{2m}{r}},
\end{align}
\end{subequations}
with $\epsilon:=-\mathrm{sgn}\big(\lambda\sin(2\lambda\kangnew)\big)$. Interestingly, the coupling $\Omega^2=\beta^n$ does not alter explicitly the equation of the mass \eqref{eq.dotm2} and it coincides with \eqref{eq.dotm}.
However, the other two equations are more involved in this case and it turns out more difficult to obtain
analytic solutions. For instance, if one tries to fix the same gauge as above
by requesting $m=m(x)$, then $N^x=0$, which leads to $\dot{\kappa}\neq 0$ and thus one still needs to solve
the two nonlinear coupled equations \eqref{eq.dotr2}--\eqref{eq.dotk2} with $N^x=0$. Instead, if one
sets $\kappa=\kappa (x)$, the conservation of such gauge condition provides a complicated form for the shift $N^x$.

In addition, it is interesting to note that the marginally bound case, defined as $\kappa=0$, is only consistent
for the trivial solution $m=0$ or if $n=0$, which is the scenario studied in the main text.
As a result, the ``marginally bound'' scenario does not exist in this model.

In fact,
despite the difficulty of finding a general solution for the system,
we have been able to obtain a particular solution that resembles the ``marginal'' case.
More precisely, for $n=1$, the explicit choice $\kappa =-2\lambdabar m/r$ is a solution to \eqref{eq.dotk2} for any choice of $m$ and $r$, and it reduces to $\kappa=0$ in the GR limit $\lambdabar\to0$. Further, if we work in the diagonal gauge
by imposing $m=m(x)$, which implies $N^x=0$, then \eqref{eq.dotr2} can be integrated explicitly:
\begin{align}\label{eq.solparticular}
    t=t_0(x) &-\frac{\epsilon{\lambdabar}}{\sqrt{1-\lambdabar}}\Bigg(2m\sqrt{\lambdabar} \log\left(\frac{\sqrt{r}-\sqrt{2\lambdabar m}}{\sqrt{r}+\sqrt{2\lambdabar m}}\right)
    +\sqrt{2 m r}\left(2+\frac{r}{3\lambdabar m}\right)\Bigg).
\end{align}
Interestingly, for this particular solution, it takes an infinite amount of dust proper time to reach the surfaces $r=2\lambdabar m$. The metric in this chart is just
\begin{align}\label{eq.metricdiag2}
    ds^{2}=&\,-{dt}^{2}+ \left(1-\frac{2\lambdabar m}{r}\right)^{-3}{(r')^{2}}{dx}^{2}+r^{2}{d\sigma}^{2},
\end{align}
with $m=m(x)$, and $r=r(t,x)$ as given in \eqref{eq.solparticular}.
As in the previous case, it can be checked that all the relevant curvature scalars are finite
at $r=2\lambdabar m$, although the metric is significantly different from \eqref{eq.metricdiag} with $n=1$.

As in the previous appendix, we make a second gauge choice $r=x$. Then, we have $N^x=\sqrt{1-\lambdabar}\sqrt{2m/r}(1-2\lambdabar m/r)$, because $\kappa=-2\lambdabar m/r$ and $n=1$. The only dynamical equation left is \eqref{eq.dotm2}, which is now determined by the above expression for the shift. The metric reads,
\begin{align}\label{eq.metricGP2}
    &d{s}^{2}=
    -\frac{r-{2m}}{r-2\lambdabar m}{dt}^{2}+2\epsilon\sqrt{1-\lambdabar}\sqrt{\frac{2m}{r}}\left(1-\frac{2\lambdabar m}{r}\right)^{-2}dtdr+\left(1-\frac{2\lambdabar m}{r}\right)^{-3}{dr}^{2}
   +r^{2}{d{\sigma}}^{2}, 
\end{align}
and just as with the previous model, this can be related to \eqref{eq.metricdiag2} through the coordinate transformation $dr=\dot{r}dt+r'dx$, with $\dot{r}=-\epsilon\sqrt{1-\lambdabar}\sqrt{2m/r}(1-2\lambdabar m/r)$.

\providecommand{\noopsort}[1]{}\providecommand{\singleletter}[1]{#1}%

\end{document}